\documentstyle[prl,aps,preprint,epsfig]{revtex}

\begin{document}

\def\xx{}
\def\eref#1{(\protect\ref{#1})}
\def\fref#1{\protect\ref{#1}}
\def\sref#1{\protect\ref{#1}}
\def\etal{{\it{}et~al.}}
\def\e{{\rm e}}
\def\d{{\rm d}}
\def\vx{{\bf x}}
\def\pstress{p_{\rm stress}}



\title{Self-organized criticality, evolution and\\
the fossil extinction record}
\author{M. E. J. Newman}
\address{Cornell Theory Center, Cornell University, Ithaca, NY 14853--3801}
\date{July 15, 1996}
\maketitle

\begin{abstract}
Statistical analysis indicates that the fossil extinction record is
compatible with a distribution of extinction events whose frequency is
related to their size by a power law with exponent $\tau\approx2$.  This
result is in agreement with predictions based on self-organized critical
models of extinction, and might well be taken as evidence of critical
behaviour in terrestrial evolution.  We argue however that there is a much
simpler explanation for the appearance of a power law in terms of
extinctions caused by stresses (either biotic or abiotic) to which species
are subjected by their environment.  We give an explicit model of this
process and discuss its properties and implications for the interpretation
of the fossil record.
\end{abstract}



\section{Introduction}
There has in the last few years been much interest in the idea that
coevolution in extended ecosystems could give rise to self-organized
critical behaviour (Kauffman~1992, Bak and Paczuski~1996, Sol\'e and
Bascompte~1996).  It has been suggested that as a result of competitive
interactions between species, ecosystems drive themselves to a critical
state in which the chance mutation of one species can spark a burst, or
``avalanche'', of evolution that can touch an arbitrary number of other
species, and potentially even the entire planet.  The attraction of this
theory is that it could provide a natural mechanism for the rapid turnover
of species seen in the fossil record without the need to invoke
environmental catastrophes to explain the extinction and replacement of
apparently well-adapted species.  Within this theory constant change is a
natural feature of evolution, and stability merely an illusion brought
about by the myopia of the observer; viewed on a sufficiently large scale
in either time or space there is nothing remotely stable about evolution.

Intriguing though this theory is, it suffers at present from a lack of hard
evidence in its favour.  There has been much discussion, and some anecdotal
evidence, of the processes which it is supposed are responsible for
self-organization at the species level (see, for example, Maynard
Smith~1989), and it has also been observed (Burlando~1990) that taxonomic
trees appear to possess a self-similar structure which might be evidence of
underlying critical processes.  For the moment however, most of our
quantitative data on terrestrial evolution come from the fossil record, a
record whose coverage and temporal resolution is sufficiently poor as to
make it difficult to distinguish between competing theories.  One hint of
an underlying complex dynamics may be the ``punctuated equilibria''
originally pointed out by Eldredge and Gould~(1972), in which species
evolve in bursts, separated by periods of stasis.  Many theories of
self-organization in evolution predict intermittent behaviour of precisely
this kind.  However the feature of the fossil record that has attracted the
most attention to date, as far as self-organization is concerned, is the
record of extinctions.

Extinction has played an important role in the evolution of life on Earth.
Of the estimated one billion or more species which have inhabited the
planet since the beginning of the Cambrian, only a few million are still
living today.  All the rest became extinct, typically within about ten
million years of their first appearance.  A variety of explanations have
been proposed for this high turnover rate.  The most traditional are that
extinction is a result of abiotic environmental stresses (Raup~1986,
Hoffmann and Parsons~1991).  For example, species can die out through
inability to evolve fast enough to keep pace with a changing
environment---a change in climate, for instance, or changes in the
chemistry of the sea.  More recently it has been suggested that some
extinction events may have been the result of sudden planet-wide
calamities, such as changes in sea-level (Hallam~1989), the impacts of
meteors or comets (Alvarez~\etal~1980), or large-scale volcanism
(Courtillot~\etal~1988, Duncan and Pyle~1988).  The alternative explanation
for extinctions is that they arise through a biotic mechanism; even though
a species may coexist in apparent stability with its competitors for
millions of years, it can still become extinct if one of those competitors
evolves to a new form, or if a new competitor expands into its territory
(Maynard Smith~1989).  Since there is evidence in support of both biotic
and abiotic mechanisms for extinction, it is probably reasonable to suppose
that both have played a part in the history of terrestrial life.

The fossil data on extinction cover about $250\>000$ fossil species, which
constitute a very small fraction of all the species that have ever lived,
but still a statistically significant sample.  Given a sufficiently large
number of fossils from any one taxon, one can infer approximate dates of
origination and extinction of that taxon, and hence, for example, measure
the distribution of lifetimes of taxa, or of the sizes of extinction
events.  It has been suggested that these data could provide a test of the
proposed self-organizing models of evolution, provided we can establish
some connection between the dynamics of evolution and the extinction rate.
The problem is that we are not even sure that such a connection exists, let
alone what it is.  A number of possibilities have been suggested recently,
based on theories of extinction caused either by species interactions, or
by environmental stress.  Bak and Paczuski~(1996) have argued that if an
apparently well-adapted species can be driven to extinction by the sudden
evolution of a competitor species, then one should expect to see heightened
extinction during periods of enhanced evolutionary activity.  If evolution
takes place in bursts, or avalanches, as predicted by self-organized
critical models, then, they argue, so might extinction.  By combining
arguments such as these with explicit mathematical models of avalanche
dynamics in evolution, like those of Kauffman and Johnsen~(1991) and Bak
and Sneppen~(1993, Sneppen~\etal~1995), it has been possible to make
predictions about the expected frequency of extinction events.  (See, for
example, Kauffman~(1995).)

Another possible connection between extinction and evolution has been
investigated in modelling work by Sol\'e~(1996).  In his model one takes a
large number of interacting species and applies a specific criterion for
deciding when the pressure placed on a species by those around it is
sufficient to drive it to extinction.  New species appear by speciation
from the survivors at precisely the rate required to keep the total number
constant.  The model displays mass-extinction events of all sizes $s$ up to
the size of the entire ecosystem.  Furthermore, if one constructs a
histogram of the relative frequency $p(s)$ of extinction events of
different sizes, the resulting curve follows a power law:
\begin{equation}
p(s) \propto s^{-\tau}.
\label{powerlaw}
\end{equation}
The power-law form is one of the most characteristic features of critical
behaviour, and has been taken as evidence of self-organization in a wide
variety of systems.  The measured exponent of the power law is
$\tau\approx2$, a prediction which should be testable against the fossil
data.  Another approach has been taken by Newman and Roberts~(1995, Roberts
and Newman~1996), who made the contrasting assumption that all extinction
events were the result of physical stresses coming from outside the
ecosystem.  In their model species underwent avalanches of coevolution but
were also subjected to environmental stresses of varying severity, which
tended to wipe out the less fit species.  This model also predicts a
power-law distribution of extinction sizes, again with exponent
$\tau\approx2$.

In fact, the power-law extinction distribution is a recurring theme of
almost all of the recent models of self-organization in evolution.  This
suggests that one might test the veracity of these models by examining the
fossil record for such power laws.  In the next section we conduct just
such an examination.  We present results which show that the data are in
fact consistent with a power law, and furthermore that we can extract quite
an accurate figure for the exponent $\tau$, which is in perfect agreement
with the more quantitative theories of extinction.  Given this satisfying
result, the next question we ask, which is addressed in the third section
of the paper, is whether this implies that terrestrial evolution is indeed
a self-organized critical process.  At the risk of giving away the
punchline, it turns out that the answer to this question is no---it implies
no such thing.  There is a much simpler and very elegant explanation for
the appearance of a power law in the extinction distribution.  We will show
that, given only the very simplest of assumptions about the causes of
extinction, we can formulate a model of the extinction process which does
not rely on coevolution for its results (although coevolution can certainly
be present) and yet predicts a power-law distribution of extinction sizes.
Our results appear to be quite independent of such niceties as the dynamics
of evolution or the difference between biotic and abiotic extinctions,
implying that we should expect to see power laws in the extinction record
whether evolution is self-organized critical or not.

\section{Does the extinction distribution follow a power law?}
\label{s1}
In this section we investigate the question of whether the available data
on the extinction of fossil species are compatible with a frequency
distribution of extinction sizes following a power law of the form given in
Equation~\eref{powerlaw}.  The data we use come from the compilation by
Sepkoski~(unpublished), as do those employed by most others who have
conducted statistical investigations of terrestrial extinction.  The
particular subset we use deals with Palaeozoic and Mesozoic marine species,
mostly invertebrates; data on marine invertebrates are far more numerous
than those for any other fossil biota.  We further restrict ourselves to
the reduced data set created by Raup~(1991), which has been edited to
remove some statistical bias (Raup and Boyajian~1988).  The edited data are
grouped into genera.  Although this grouping has the effect of reducing the
number of extinctions in the data set, it has the advantage of increasing
the precision with which we know the date of any particular extinction.  A
summary of the data set is given in Table~A1 of Raup~(1991).

\begin{figure}
\begin{center}
\psfig{figure=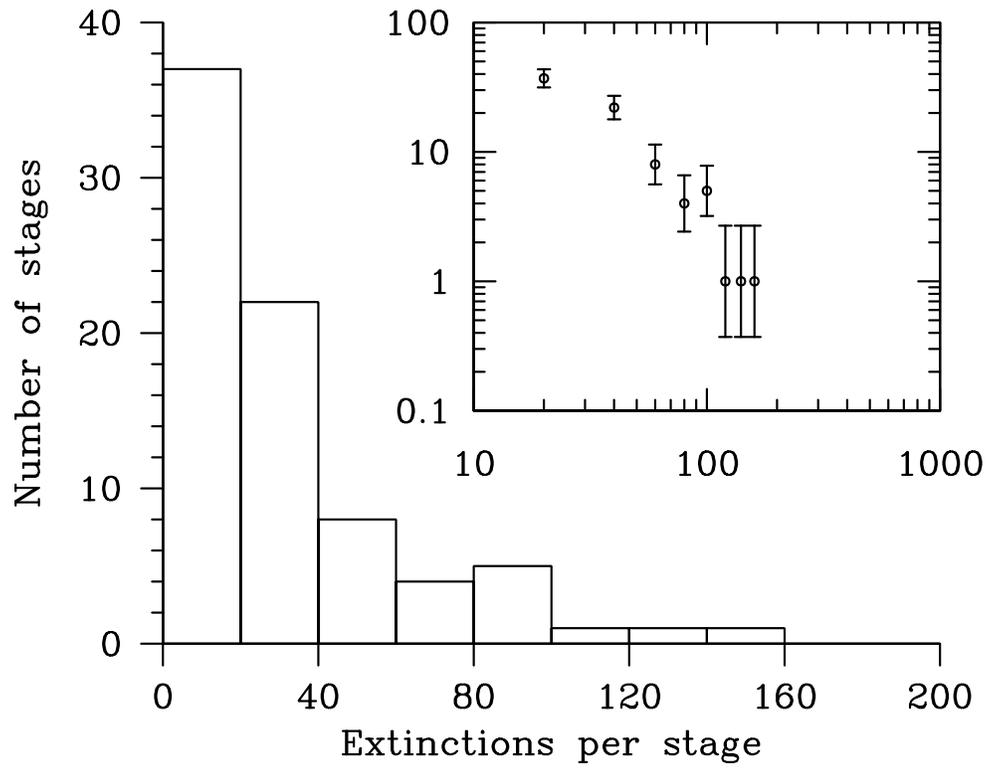,width=13cm}
\end{center}
\caption{Histogram of the relative frequency of extinction events of
various sizes on linear scales and (inset) on logarithmic ones.
\label{histogram}}
\end{figure}

The most straightforward thing to do with the data is to divide the time
they span into intervals, count the number of species becoming extinct in
each interval (which is a measure of extinction rate) and make a histogram
of the results.  As discussed by Raup~(1986), the divisions of time chosen
for such analyses are usually the stratigraphic stages.  The results of the
histogramming are shown both on linear and on logarithmic scales in
Figure~\fref{histogram}.  (On the latter the power-law distribution should
appear as a straight line.)  This oft-reproduced plot has been used by a
variety of commentators as evidence both of the existence and the
non-existence of a power-law extinction distribution in the fossil record.
However, the errors on the histogram, shown as bars in the logarithmic
plot, are really too large to allow the question to be answered one way or
the other.  To get an answer we must resort to statistical analysis of a
more sophisticated nature.

Although not originally intended to address this particular issue, a
suitable method of analysis has been given by Raup~(1991), who introduced
the concept of the ``kill curve'' and a technique for deducing it from the
fossil record by comparing Monte Carlo calculations of genus survivorship
with fossil survivorship data.  It turns out that the kill curve is closely
related to the distribution of extinctions.  Let us look first at the
original calculations performed by Raup.

\begin{figure}
\begin{center}
\psfig{figure=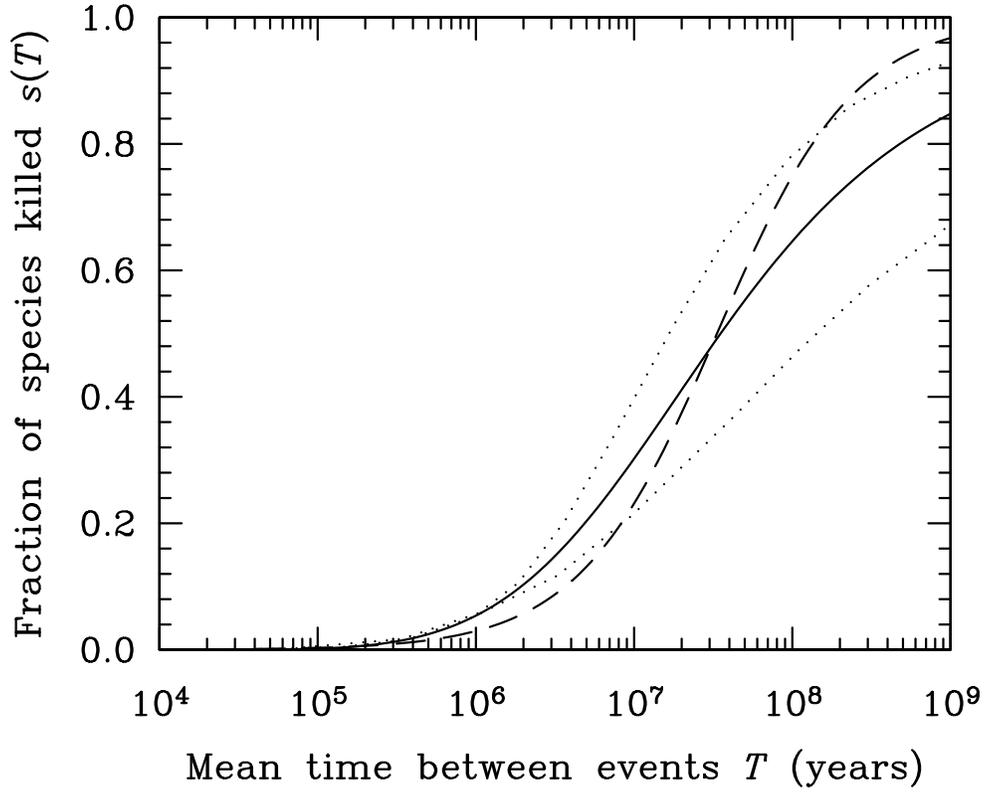,width=13cm}
\end{center}
\caption{Kill curves for Phanerozoic marine species.  The solid line is the
curve given by Raup~(1991) and the dotted lines are the associated errors.
The dashed line is the best fit kill curve of the form given in
Equation~\eref{newmanform}.
\label{killcurve}}
\end{figure}

Consider Figure~\fref{killcurve}.  The solid line shows Raup's kill curve.
The curve is a cumulative frequency distribution of extinctions.  The
horizontal axis of the plot tells you how long a time $T$ you should expect to
wait on average between extinction events which kill a fraction $s$ or more of
the species in the ecosystem, where $s$ is given on the vertical axis.  Note
that $s$ is measured as a fraction of the total number of species, not genera,
in the system.  Although genus-level data are used to calculate the curve, it
is extrapolated down to the species level, since this is of more relevance to
biological issues.  The curve in the figure was found by assuming the
functional form
 \begin{equation}
 s(T) = {[\log T]^a\over\e^b + [\log T]^a},
 \label{raupform}
 \end{equation}
 and then fitting the parameters $a$ and $b$ to the fossil data using a method
similar to the one described below.  As Raup emphasizes, the form of
Equation~\eref{raupform} is not based on any particular theory of extinction.
It was merely chosen because it has the expected sigmoidal form and is
reasonably flexible.  Using Sepkoski's data, Raup finds a best fit to the
fossil record when $a=5$ and $b=10.5$.  These figures give the solid curve
in Figure~\fref{killcurve}, whilst the dotted ones give the estimated error.

\begin{figure}
\begin{center}
\psfig{figure=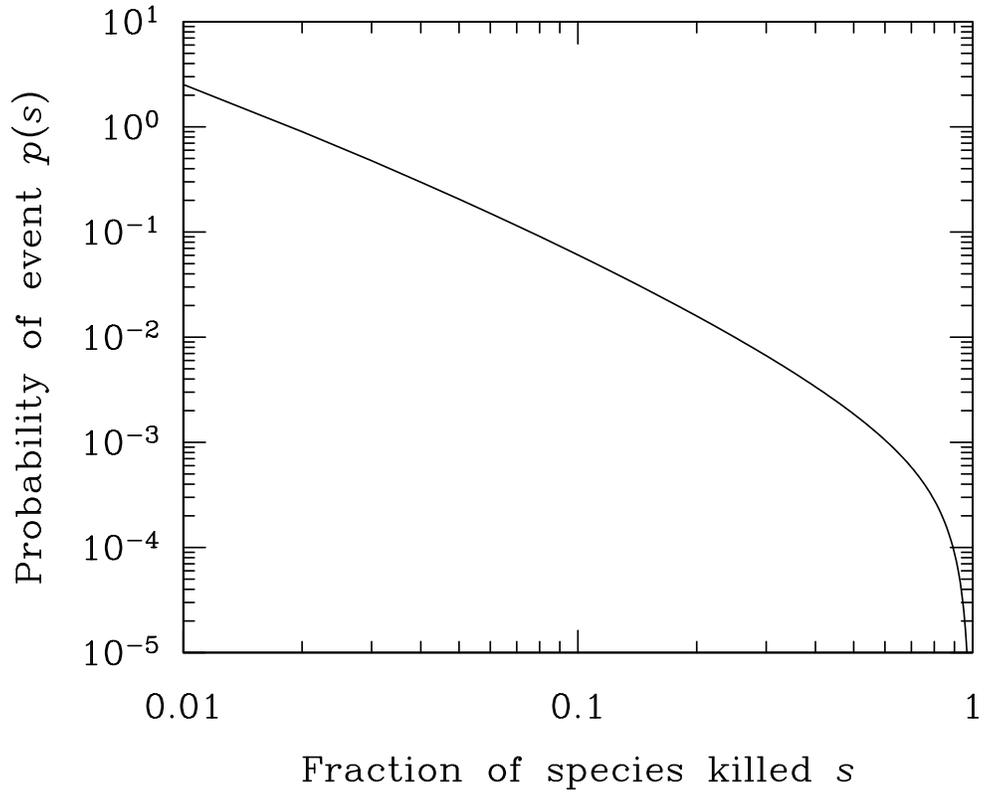,width=13cm}
\end{center}
\caption{The extinction distribution corresponding to the solid kill curve
in Figure~\fref{killcurve}.
\label{raupext}}
\end{figure}

We now ask, what is the connection between the kill curve and the
distribution of extinctions?  Let us denote by $p(s)\>\d s$ the number of
extinctions with size between $s$ and $s+\d s$ taking place per unit time.
The function $p(s)$ is precisely the probability distribution of
extinctions that we wish to calculate.  The number of extinctions $P(s)$ of
size {\em greater\/} than $s$ per unit time is
\begin{equation}
P(s) = \int_s^1 p(s') \>\d s'.
\label{Ps}
\end{equation}
The mean time $T$ between events of size $s$ or greater is just $1/P(s)$,
and thus we establish a connection between $s(T)$ and $p(s)$.  Making use
of this connection, we have plotted in Figure~\fref{raupext} the extinction
distribution $p(s)$ corresponding to Raup's kill curve, and on logarithmic
scales the result is something approximating a straight line, although it
falls off as the size of the extinctions approaches unity.  (A similar
fall-off is seen in most theoretical models of extinction, so this should
not be regarded as a grave problem.)  We can conclude therefore that the
published results of Raup imply an extinction distribution which
approximately follows a power law, and by fitting a straight line to
Figure~\fref{raupext} we can extract a figure of $\tau = 1.9 \pm 0.4$ for
the exponent of the power law.

Taking the analysis a step further, we can also ask what happens if we {\em
assume\/} a power-law form for the extinction distribution, calculate the
corresponding kill curve and fit that to the fossil data.  Reversing the
argument given above, it is not hard to show that the form for the kill
curve corresponding to a power-law is
\begin{equation}
s(T) = \biggl[ {T_0\over T} + 1 \biggr]^{1\over1-\tau}.
\label{newmanform}
\end{equation}
Like Raup's kill curve, this one has two free parameters: the exponent
$\tau$, and a ``typical waiting time'' $T_0$.  We fit these parameters to
the fossil data as follows.

First we note that the two parameters are not independent.  The mean
extinction rate per species for the period covered by the data set is known
to be about one species per 4 million years (Raup~1991), and this is
related to the extinction distribution by
\begin{equation}
\mbox{mean extinction rate} = \int s\,p(s)\>\d s.
\end{equation}
This gives us a constraint on the values of $\tau$ and $T_0$ such that if
we know $\tau$, the value of $T_0$ is fixed.  As a result there is
essentially only one parameter in the problem to be fitted to the data, and
that is the exponent $\tau$.

The fitting is done by Monte Carlo simulation of genus survivorship.  In
this simulation genera are founded by a single species and for each unit of
succeeding time there is a constant probability of speciation for each
species in the genus, and a random time-varying probability of extinction
which is distributed according to the kill curve,
Equation~\eref{newmanform}, or equivalently according to the extinction
distribution, Equation~\eref{powerlaw}.  One performs the simulation many
times, starting each time with a large number of genera, and continuing
until all of them become extinct.  Then one plots the fraction of surviving
genera as a function of time for each simulation, giving a set of
survivorship curves.  For a model with stochastically constant extinction,
one expects all these curves to be identical, except for statistical
variation due to the finite number of genera taking part.  However, with an
extinction rate which varies over time, as we are here assuming, we expect
to see an intrinsic dispersion in this set of curves, and it is this
dispersion that we compare with the fossil data.  It is by essentially this
method that Raup extracted values for the parameters $a$ and $b$ appearing
in Equation~\eref{raupform}.  We have performed the calculation using the
kill curve described by Equation~\eref{newmanform}, and find that the
survivorship curves extracted agree as well with the fossil data as those
presented by Raup~(1991), and hence we can conclude that the data are
indeed compatible with a power law extinction distribution.  Furthermore,
under the assumption that the distribution is a power law, we can extract a
value for the exponent $\tau$ of the power law.  That value is:
\begin{equation}
\tau = 2.0 \pm 0.2.
\label{tauvalue}
\end{equation}
This figure is in good agreement with figures for the same exponent from
the models of Newman and Roberts~(1995) and of Sol\'e~(1996).  The
corresponding kill curve is shown as the dashed line in
Figure~\fref{killcurve}, and agrees with the curve given by Raup, within
the quoted accuracy.

So, if the fossil extinction distribution does follow a power law, can we
take this as evidence that the underlying evolutionary dynamics is of a
self-organized critical nature?  This is the question we address in the
next section.

\section{Why does the extinction distribution follow a power law?}
\label{s2}
We have shown that the fossil data for marine species compiled by Sepkoski
are consistent with the existence of a power-law distribution of extinction
sizes with an exponent close to 2.  In the light of the arguments reviewed
in Section~I, would we then be justified in claiming, tentatively perhaps,
to have found evidence that terrestrial evolution is a self-organized
critical phenomenon?  In this section we argue that this is not a justified
assumption---that there is in fact a much simpler explanation for the
appearance of a power law.  Using a minimum of assumptions about the nature
of evolution and extinction, we formulate a new model which predicts that
the extinction distribution should follow a power law regardless of the
nature of the underlying evolutionary processes, and indeed regardless of
the precise causes of extinction.  This model offers at once an elegant
explanation of the observed extinction distribution, and at the same time a
(perhaps slightly disheartening) demonstration that we cannot hope to learn
much about the nature of the evolutionary process by examining this
distribution.

The principal assumption of our new model is that the extinction of a
species is caused by a change in its environment.  Changes may take many
forms, and in particular may be either biotic or abiotic in nature: they
may be due either to changes in the other species with which a species
interacts, or to environmental effects such as climate change.  Let us call
these causes of extinction ``stresses''.  During its lifetime, a species
will in general experience a number of these stresses, and for any given
stress each species will have a certain tolerance (or lack of it).  We
quantify this tolerance in our model by a species fitness measure which we
denote $x$.  When the stress occurs, species with higher values of $x$ are
less likely to become extinct than those with lower values.  The ability to
withstand stress could depend on many factors, such as ability to adapt to
a new environment (``generalists'' vs.\ ``specialists''), or possession of
particular physical attributes.  (Large body mass, for example, appears to
have been a disadvantage during the extinction event which ended the
Cretaceous.)  For our purposes however, we will not need to know exactly
what properties the fitness depends on; all that will matter for us is that
such a fitness can be defined.  For convenience, $x$ is assumed to take
values between zero and one, though this is not a necessary condition for
any of the results discussed later.

We also need to define a measure of the size or strength of our stresses,
so as to distinguish between, for example, the impacts of large and small
meteors on the Earth.  In our model we therefore divide time up into short
intervals, and define another number, denoted $\eta$, which measures the
level of stress during a given interval.  When the stress level is high we
expect many species to become extinct in that time interval; when it is
low, few will become extinct.  We have chosen a simple rule for our model
to achieve this result: if at any time the numerical value of $\eta$
exceeds that of the fitness $x$ of a species, that species becomes extinct.
One may well worry about how we choose the values of the stress level
$\eta$.  However, as we will show, the result that the extinction
distribution is a power law does not depend on what choice we make (though
the precise exponent of the power law does).  Our only assumptions will be
that small stresses are more common than large ones, and that, based on the
evidence of terrestrial prehistory, stresses large enough to wipe out every
species on the planet are uncommon.

In order to complete our model there are a couple of other components we
need to add.  First, we assume that the number of species our ecosystem can
sustain is roughly constant over time.  To satisfy this constraint, we
introduce after every extinction event new species equal in number to those
that have become extinct.  We need to choose values for the fitnesses of
these new species.  The two obvious ways to choose them would be either by
``inheriting'' fitnesses from survivor species (from which they are assumed
to have speciated), or by giving them purely random fitness values.  Again,
it turns out that the qualitative predictions of the model are independent
of the exact choice we make.

Finally, we observe that, if this were all there was to our model, its
dynamics would soon come to a standstill, when all the species with low
fitnesses had been eliminated and all those remaining were susceptible only
to very large stresses, which, as we have said, are rare.  Clearly this
does not happen in a real ecosystem, and the explanation is clear:
evolution.  In the intervals of time between large stresses on the
ecosystem, the selection pressure of the stress is not felt and species
evolve under other competing pressures, possibly at the expense of their
ability to survive stress.  Thus, over time, species' fitnesses with
respect to stress may increase or decrease.  Again there are two
contrasting views about how this might take place.  The gradualist
viewpoint would be that fitnesses wander slowly and continually over time.
The alternative is the punctuated equilibrium viewpoint, under which, in
any given interval of time, some small fraction $f$ of species would evolve
to radically new forms, assuming completely different values for their
fitnesses.  Once more, it turns out that the fundamental predictions of our
model do not depend on which choice we make.

\begin{figure}
\begin{center}
\psfig{figure=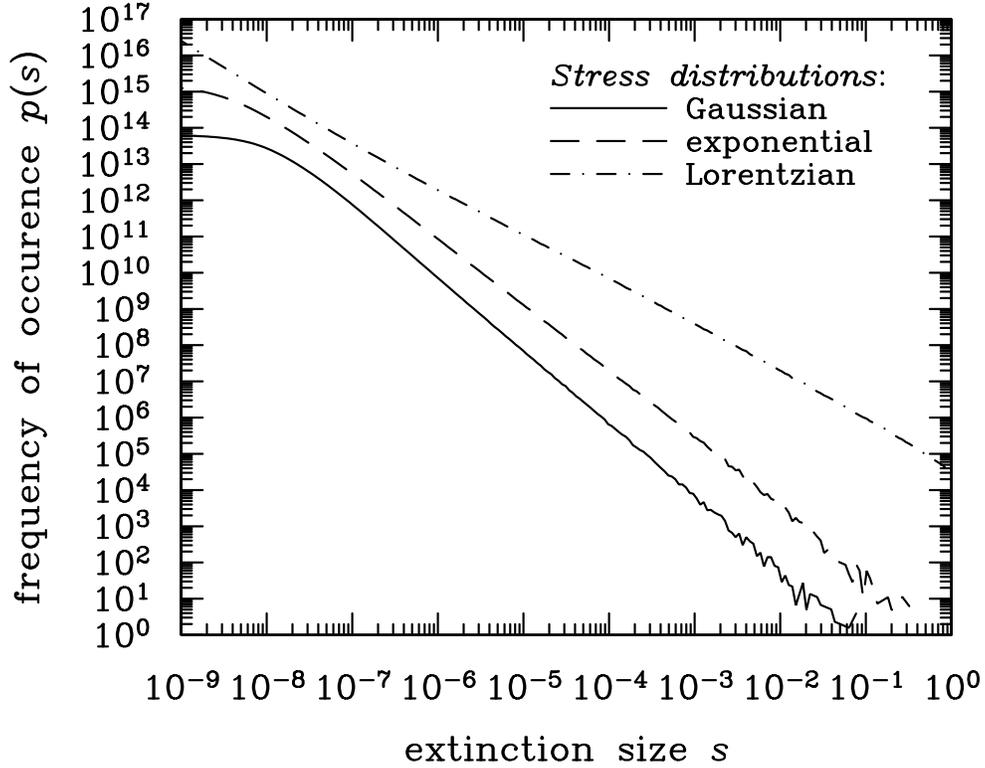,width=13cm}
\end{center}
\caption{The distribution of extinction sizes measured in simulations of
the model for three different forms of the stress distribution
$\pstress(\eta)$.  The exponents of the power laws are $\tau=2.02\pm0.02$
for the Gaussian, $\tau=1.84\pm0.03$ for the exponential, and
$\tau=1.24\pm0.04$ for the Lorentzian.
\label{extinctions}}
\end{figure}

We are led then to a new model of extinction which in its simplest form is
as follows.  We have an ecosystem consisting of some large number $N$ of
species.  With each species $i$ we associate a fitness $x_i$ which can take
values between zero and one.  We then execute the following steps
repeatedly.
\begin{enumerate}
\item At each time-step, a small fraction $f$ of the species, selected at
random, evolve, and their fitnesses $x_i$ are changed to new values chosen
at random in the range $0\le x_i<1$.
\item We choose a stress level $\eta$ randomly from some distribution
$\pstress(\eta)$, and all species whose fitnesses $x_i$ lie below that
value become extinct and are replaced by new species whose fitnesses are
also chosen at random in the range $0\le x_i<1$.
\end{enumerate}
Many variations are possible.  We discuss some of the more important ones
below, but for the moment let us examine the predictions of this version of
the model.  We have performed extensive simulations of the model using a
wide variety of different choices for the form of the stress distribution
$\pstress(\eta)$, including forms with a power-law fall-off away from zero
(such as a Lorentzian) which might be expected if the stresses were
primarily due to coevolutionary avalanches, and forms with an exponential
fall-off (such as an exponential or a Gaussian) which might be more
appropriate for abiotic stresses.  The results are shown in
Figure~\fref{extinctions}.  In each case the distribution of extinctions
closely follows a power law over a wide range of sizes $s$, deviating only
at very small values of $s$.  The exponent of the power law depends on the
exact form of the stress, but its existence does not.  There appear to be
only two conditions for producing a power law, and they are (i)~that the
fraction $f$ of species evolving in each time-step should be small $f\ll
1$, but non-zero (in order that the dynamics does not grind to a halt, as
described above), and (ii)~that the chances of getting a stress of
sufficient magnitude to wipe out every species in the system should be very
small.  If these conditions are satisfied, then in every case we find a
power-law distribution of extinction sizes.

In this simplest form, the model is in fact exactly the same as a model
used by Newman and Sneppen~(1996) to model the dynamics of earthquakes.
Their paper gives a detailed mathematical analysis of the model, explaining
the appearance of the power law within a time-averaged approximation and
also giving an explanation of the conditions (i) and (ii) above.  Rather
than reproduce that discussion here, we refer the reader to that paper, and
here discuss instead the connection of our model to real extinction, an
issue which brings up a number of important questions.  First, there are
the questions of the form of the stress distribution $\pstress(\eta)$, the
choice of the fitness for newly appearing species, and the particular
dynamics we have chosen to represent the evolution process.  Depending on
which school of thought he or she adheres to, the reader might well have
chosen these features of the model differently.  However, as we have
already mentioned, such changes have no effect on the appearance of a power
law extinction distribution.  This fact is depicted explicitly in
Figure~\fref{extinctions}, in which the extinction distributions for
various choices of $\pstress(\eta)$ are compared.  A more important
objection to the model is that we have assumed that every stress on the
system affects every species.  This is clearly not realistic.  Some
stresses will for example be localized in space, or will not reach under
the sea, or will only reach under the sea, and so forth.  This kind of
situation can be accounted for by considering a different fitness function
$x$ for such stresses, one for example which is much higher if you live
under the sea, or in Africa, or whatever.  But this now means that we have
two (or more) fitness functions for each species.  This brings us on to
another important issue, which is that there are of course many types of
stress.  Not all stresses are equivalent, and some may affect certain
species more than others.  Thus it is inadequate to have just one fitness
function for each species.  We should instead have many such
functions---one for each type of stress.  This leads us to a more
sophisticated ``multi-trait'' version of the model in which each species
has a number $M$ of different fitness functions, or equivalently an
$M$-component vector fitness $\vx$, and $M$ corresponding different types
of stress.  The dynamics of the model are exactly as before, except that
now a species becomes extinct if {\em any one\/} of the types of stress to
which it is subject exceeds its threshold for withstanding that type of
stress.  For all values of $M$ which we have investigated (up to $M=50$),
this model too shows power-law distributions of extinction sizes.

It thus appears that for both the simplest version of our model, and for
all reasonable variants, the power law distribution of extinction sizes is
an inevitable result.  The only required features are stresses, either
biotic or abiotic to which species are subjected, and varying abilities of
the species to withstand these stresses.  Nothing about this model requires
that we have a self-organized critical evolutionary dynamics taking place;
coevolutionary avalanches, of the kind supposed by earlier authors to be
responsible for power-law behaviour in the extinction size distribution,
are not a component of our model.

\section{Conclusions}
Based on modelling work by a number of authors, the claim has been made
that a self-organized critical evolutionary dynamics should give rise to a
power-law distribution of extinction sizes.  In this paper we have first
investigated the nature of the fossil record and concluded that it is
entirely compatible with such an extinction distribution.  Second, we have
asked whether this implies that evolution is indeed a critical process, a
question which we have answered by proposing a new and simple model for
extinction which makes no assumptions about the dynamics of the evolution
process but which nonetheless predicts a power-law extinction distribution
in every case we have investigated.  We thus conclude that there is no
evidence in the distribution of fossil extinction events to support the
notion of self-organized critical behaviour in evolution.  What we have
instead, is a rather elegant picture of an empirical result---the power-law
distribution of extinctions in the fossil record---and its explanation in
terms of a theory of extinction caused by stresses on the ecosystem.

\section{Acknowledgements}
The author is very grateful to David Raup for his help with the
calculations described in Section~\sref{s1}, and to Kim Sneppen for many
fruitful discussions concerning the model presented in Section~\sref{s2}.
This work was funded in part by the NSF under grant number ASC-9404936, and
by the Cornell Theory Center.

\vspace{1cm}

\begin{list}{}{\leftmargin=2em \itemindent=-\leftmargin%
\itemsep=0pt \parsep=0pt}

\item {\frenchspacing Alvarez, L. W., Alvarez, W., Asara, F., and Michel,
H. V.  1980 Extra-terrestrial cause for the Cretaceous/Tertiary extinction.
{\it Science\/} {\bf208}, 1095.}

\item {\frenchspacing Bak, P. and Paczuski, M. 1996 Mass extinctions
vs. uniformitarianism in biological evolution.  Brookhaven National
Laboratory preprint no. 62678.}

\item {\frenchspacing Bak, P. and Sneppen, K. 1993 Punctuated equilibrium
and criticality in a simple model of evolution.  {\it Phys. Rev. Lett.}
{\bf71}, 4083.}

\item {\frenchspacing Burlando, B. 1990 The fractal dimension of taxonomic
systems.  {\it J. Theor. Biol.} {\bf146}, 99.}

\item {\frenchspacing Courtillot, V., F\'eraud, G., Maluski, H., Vandamme,
D., Moreau, M. G. and Besse, J. 1988 Deccan flood basalts and the
Cretaceous/Tertiary boundary.  {\it Nature\/} {\bf333}, 843.}

\item {\frenchspacing Duncan, R. A., and Pyle, D. G. 1988 Rapid eruption of
the Deccan flood basalts at the Cretaceous/Tertiary boundary.  {\it
Nature\/} {\bf333}, 841.}

\item {\frenchspacing Eldredge, N. and Gould, S. J. 1972 Punctuated
equilibria: an alternative to phyletic gradualism.  In {\it Models in
Paleobiology\/} T. J. M. Schopf (Ed.), Freeman, San Francisco.}

\item {\frenchspacing Hallam, A. 1989 The case for sea-level change as a
dominant causal factor in mass extinction of marine invertebrates.  {\it
Phil. Trans. Roy. Soc. B\/} {\bf325}, 437.}

\item {\frenchspacing Hoffmann A. A. and Parsons, P. A. 1991 {\it
Evolutionary Genetics and Environmental Stress,} OUP, Oxford.}

\item {\frenchspacing Kauffman, S. A. 1992 {\it The Origins of Order,} OUP,
Oxford.}

\item {\frenchspacing Kauffman, S. A. 1995 {\it At Home in the Universe,}
OUP, Oxford.}

\item {\frenchspacing Kauffman, S. A. and Johnsen, S. 1991 Coevolution
to the edge of chaos: Coupled fitness landscapes, poised states, and
coevolutionary avalanches.  {\it J. Theor. Biol.}  {\bf149}, 467.}

\item {\frenchspacing Maynard Smith, J. 1989 The causes of extinction.  {\it
Phil. Trans. Roy. Soc. B\/} {\bf325}, 241.}

\item {\frenchspacing Newman, M. E. J. and Roberts, B. W. 1995 Mass
extinction: Evolution and the effects of external influences on unfit
species.  {\it Proc. Roy. Soc. B\/} {\bf260}, 31.}

\item {\frenchspacing Newman, M. E. J. and Sneppen, K. 1996 Avalanches,
scaling, and coherent noise.  Cornell Theory Center preprint
no. 96TR237.}

\item {\frenchspacing Raup, D. M. 1986 Biological extinction in Earth
history.  {\it Science\/} {\bf231}, 1528.}

\item {\frenchspacing Raup, D. M. 1991 A kill curve for Phanerozoic
marine species.  {\it Paleobiology\/} {\bf17}, 37.}

\item {\frenchspacing Raup, D. M. and Boyajian, G. E. 1988 Patterns of
generic extinction in the fossil record.  {\it Paleobiology\/} {\bf14}, 109.}

\item {\frenchspacing Roberts, B. W. and Newman, M. E. J. 1996 A model for
evolution and extinction.  {\it J. Theor. Biol.} {\bf180}, 39.}

\item {\frenchspacing Sneppen, K., Bak, P., Flyvbjerg, H. and Jensen,
M. H. 1995 Evolution as a self-organized critical phenomenon.  {\it
Proc. Nat. Acad. Sci.} {\bf92}, 5209.}

\item {\frenchspacing Sol\'e, R. V. 1996 On macroevolution, extinctions and
critical phenomena.  {\it Complexity,} {\bf1}, No. 6, 40.}

\item {\frenchspacing Sol\'e, R. V. and Bascompte, J. 1996 Are critical
phenomena relevant to large-scale evolution?  {\it Proc. Roy. Soc. B\/}
{\bf263}, 161.}

\end{list}

\newpage

\newpage

\newpage

\end{document}